\documentclass[pra,onecolumn]{revtex4}
\usepackage{epsfig}
\usepackage{graphicx}
\usepackage{amssymb}

\newtheorem{corollary}{Corollary}
\newtheorem{theorem}{Theorem}
\newtheorem{lemma}{Lemma}

\newcommand{\Tr}{\textrm{Tr}}	

\begin{document}

\title{Optimal unambiguous state discrimination of two density matrices: A second class of exact solutions}
\author{Philippe Raynal$^1$}
\author{Norbert L\"utkenhaus$^{1,2}$}
\affiliation{$^1$Quantum Information Theory Group, Institut f\"ur Theoretische Physik I, and \\
Max-Plank-Forschungsgruppe, Institut f\"ur Optik, Information und Photonik\\ Universit\"at
Erlangen-N\"urnberg, Staudtstr. 7, D-91058 Erlangen, Germany}
\affiliation{$^2$Institute of Quantum Computing and Department of Physics and Astronomy\\
University of Waterloo, Waterloo, Ontario, N2L 3G1, Canada}
\date{\today}

\begin{abstract}
We consider the Unambiguous State Discrimination (USD) of two mixed quantum states. We study the rank and the spectrum of the elements of an optimal USD measurement. This naturally leads to a partial fourth reduction theorem. This theorem shows that either the failure probability equals its overall lower bound given in terms of the fidelity or a two-dimensional subspace can be split off from the original Hilbert space. We then use this partial reduction theorem to derive the optimal solution for any two equally probable Geometrically Uniform (GU) states $\rho_0$ and $\rho_1=U\rho_0 U^\dagger$, $U^2={\openone}$, in a four-dimensional Hilbert space. This represents a second class of analytical solutions for USD problems that cannot be reduced to some pure state cases. We apply our result to answer two questions that are relevant in implementations of the Bennett and Brassard 1984 quantum key distribution protocol using weak coherent states.
\end{abstract}

\maketitle

\section{Introduction}
Quantum State Discrimination is a crucial task in Quantum Information Theory, especially in a communication context. Whenever the signal states are non-orthogonal quantum states, perfect discrimination becomes impossible and one must resort to various discrimination strategies. One might, for example, consider an error-free discrimination of the states. In that case, due to the non-orthogonality of the signal states, the measurement will sometimes fail to identify conclusively the signal. The goal therefore is to minimize the probability of inconclusive results, the so-called failure probability. This strategy is known as {\em Unambiguous State Discrimination} (USD).\\
\indent The problem of unambiguously discriminating pure states with equal {\it a priori} probabilities was solved by Dieks \cite{dieks88a}, Ivanovic \cite{ivanovic87a}, and Peres \cite{peres88a}. Later Jaeger and Shimony presented the general solution for two pure states with arbitrary {\it a priori} probabilities \cite{jaeger95a}. Shortly after this result, Chefles proved that only linearly independent pure states can be unambiguously discriminated \cite{chefles98b}. Chefles and Barnett then provided the optimal failure probability and its corresponding optimal measurement for $n$ symmetric states \cite{chefles98a}. Finally Sun {\it et al} \cite{sun02b} showed that the unambiguous discrimination of pure states is a convex optimization problem \cite{vandenberghe96a,vandenberghe04,bental01}. This result was later extended to mixed states by Eldar \cite{eldar03b}. With respect to USD of mixed states, three reduction theorems related to simple geometrical considerations were developed \cite{raynal03a,raynal05a,raynal06a}. They allow us to reduce USD problems to simpler ones where a solution might be known. Important examples of reducible problems are {\em Unambiguous State Discrimination of two mixed states with one-dimensional kernel} {\cite{rudolph03a}}, {\em Unambiguous Comparison of two pure states} \cite{barnett03a,kleinmann05,herzog05a}, {\em Unambiguous Comparison of $n$ pure states with equal {\it a priori} probabilities and equal and real overlaps} \cite{raynal06a}, {\em State Filtering} \cite{sun02a,bergou03a,herzog05b} and {\em Unambiguous Discrimination of two subspaces} \cite{bergou06}. The three reduction theorems also define a {\it standard} form of USD problems. Such a standard form corresponds to the unambiguous discrimination of two density matrices of rank $r$ in a $2r$-dimensional Hilbert space without trivial orthogonal subspace and without block diagonal structure \cite{raynal03a,raynal05a,raynal06a}. Interestingly, simple conditions to identify 2x2 block diagonal structures were derived in \cite {kleinmann07a}. When a USD problem is not of the standard form, it can immediately be reduced to simpler ones. Necessary and sufficient conditions for a USD measurement to be optimal were derived by Eldar \cite{eldar04a}. In addition lower and upper bounds on the failure probability  were provided \cite{rudolph03a, guo06a}. Recently, a first class of exact solutions was found {\cite{raynal05a}. This class corresponds to pairs of mixed states such that the lower bound on the failure probability can be reached. A subclass was later studied in \cite{herzog06a}.\\
\indent In this paper we first study the rank of the elements of an optimal USD measurement. We provide a theorem that reveals constraints on the rank of the elements associated to a conclusive detection. If we consider two density matrices $\rho_0$ and $\rho_1$ together with their respective {\it a priori} probabilities $\eta_0$ and $\eta_1$, these constraints depend on the positivity of the two operators $\rho_0-\sqrt{\frac{\eta_1}{\eta_0}} \sqrt{\sqrt{\rho_0}\rho_1\sqrt{\rho_0}}$ and $\rho_1-\sqrt{\frac{\eta_0}{\eta_1}}\sqrt{\sqrt{\rho_1}\rho_0\sqrt{\rho_1}}$. Note that the positivity of these two operators was already introduced in \cite{raynal05a} as a necessary and sufficient condition for the failure probability to reach the overall lower bound $2\sqrt{\eta_0\eta_1}\Tr(\sqrt{\sqrt{\rho_0}\rho_1\sqrt{\rho_0}})$.\\
\indent A corollary to our first theorem can be derived assuming a standard USD problem. If the two operators $\rho_0-\sqrt{\frac{\eta_1}{\eta_0}} \sqrt{\sqrt{\rho_0}\rho_1\sqrt{\rho_0}}$ and $\rho_1-\sqrt{\frac{\eta_0}{\eta_1}}\sqrt{\sqrt{\rho_1}\rho_0\sqrt{\rho_1}}$ are not positive semi-definite, then a two-dimensional subspace can be split off from the original USD problem.\\
\indent The main result of the present paper is a consequence of our first theorems. We give a second class of exact solutions for generic USD problems. This class corresponds to any pair of equally probable Geometrically Uniform (GU) states in four dimensions. Two GU states are two unitary similar density matrices $\rho_0$ and $\rho_1=U\rho_0 U^\dagger$ where the unitary matrix $U$ is an involution i.e.\ $U^2={\openone}$ . We find that only two options exist depending on the positive semi-definiteness of the operator $\rho_0- \sqrt{\sqrt{\rho_0}\rho_1\sqrt{\rho_0}}$. For these two cases we provide the optimal failure probability as well as the optimal measurement. We then apply our result to answer two relevant questions related to the implementation of the Bennett and Brassard 1984 cryptographic protocol. First 'With what probability can an eavesdropper unambiguously distinguish the {\it basis} of the signal?' and second 'With what probability can an eavesdropper unambiguously determine which {\it bit value} is sent without being interested in the knowledge of the basis?'.\\
\indent This paper is organized as follows. In Section~\ref{rank} we derive a theorem about the rank of the elements of an optimal USD measurement and give a corollary which takes the form of a reduction theorem. In Section~\ref{class} we present the exact solution for unambiguously discriminating two GU states in a four-dimensional Hilbert space. In Section~\ref{BB84} we consider two examples of practical interest for quantum cryptography. 
We then conclude in Section~\ref{conclusion}.

\section{Rank and spectrum of the elements of an optimal USD measurement}\label{rank}

In Unambiguous State Discrimination the signal states must be identified without error. If those signal states are non-orthogonal quantum states, no perfect discrimination is possible and any USD measurement will lead to some inconclusive result. The rate of such inconclusive results is called the failure probability. Given a set of signal states $\rho_i$ together with their {\it a priori} probabilities $\eta_i$ we want to find an optimal measurement that minimizes the failure probability. The measurement is a generalized measurement, that is, a set of Hermitian and positive semidefinite operators (called a positive operator-valued measure or POVM \cite{helstrom76a}) $\{E_k\}_k$ that add up to the identity, i.e.\ $\sum_k E_k=\openone$. Given a set of $N$ signal states, we consider measurements with $N+1$ outcomes, where $N$ outcomes identify conclusively one and only one signal state while the last outcome indicates that the identification failed. The $N+1$ POVM elements are denoted by $E_k$, $k=1,\dots,N$, and $E_?$, respectively.  The probability to obtain the outcome $E$ for a state $\rho$ is given by the trace quantity $\Tr(E \rho)$. Therefore the conditions for an error-free measurement  simply is $\Tr(E_k \rho_i)=0$ whenever $k\neq i$ so that only the state $\rho_k$ can trigger the measurement outcome $E_k$. The failure probability $Q$ to be optimized over all possible USD measurements is given by $Q=\sum_i \eta_i \Tr (E_? \rho_i)$. We often note $Q_i$ the partial failure probability $\eta_i \Tr (E_? \rho_i)$.\\
In this paper we consider the unambiguous discrimination of two signal states $\rho_0$ and $\rho_1$ with {\it a priori} probabilities $\eta_0$ and $\eta_1$. Consequently our measurement contains three elements $\{E_0,E_1,E_?\}$ which correspond respectively to the conclusive detection of $\rho_0$, to the conclusive detection of $\rho_1$ and to an inconclusive result. For any Hermitian and positive semi-definite operator $A$, we can introduce the notions of support, kernel and square root. The support ${\cal S}_A$ of $A$ is the subspace spanned by the eigenvectors of $A$ (eigenvectors associated with non zero eigenvalues). Its orthogonal complement is called the kernel of $A$ and is denoted ${\cal K}_A$. The square root $\sqrt{A}$ of $A$ is defined as the unique positive semi-definite operator such that $\sqrt{A}^2=A$. This square root operator allows us to write decompositions of the form
\begin{eqnarray}\label{dec}
A=MM^\dagger \,\, {\rm with} \,\, M=\sqrt{A}V,
\end{eqnarray}
for any unitary transformation $V$. Since the states $\rho_i$, $i=0,1$ and the POVM elements $E_k$, $k=0,1,?$ are positive semi-definite operators, we can introduce their support, kernel, square root, and decompositions of the form given in Eqn.(\ref{dec}).\\

We now derive upper bounds on the rank of the elements $E_0$ and $E_1$ of an optimal USD POVM. The object of our first theorem will be to find under which conditions these upper bounds can be reached. To start, we recall that the error-free condition $\Tr[E_i \rho_j]=0$, $i,j=0,1$, $i \neq j$, implies the orthogonality between the support of $E_i$ and the support of $\rho_j$ \cite{raynal03a}. Therefore ${\cal S}_{E_i} \subset {\cal K}_{\rho_j}$ and the dimension of the support ${\cal S}_{E_i}$, or equivalently the rank $r_{E_i}$ of $E_i$, is upper bounded by the dimension of the kernel ${\cal K}_{\rho_j}$:
\begin{eqnarray}
r_{E_0}&\le&dim({\cal K}_{\rho_1}), \label{first}\\ 
r_{E_1}&\le&dim({\cal K}_{\rho_0}). \label{second}
\end{eqnarray}
In the relevant case of two density matrices without overlapping supports, $dim({\cal H})=dim({\cal S}_{\rho_0})+ dim({\cal S}_{\rho_1})$ and Eqn.(\ref{first}) and (\ref{second}) simplify to:
\begin{eqnarray}\label{r0r1}
r_{E_0} &\le& r_0, \\
r_{E_1} &\le& r_1,
\end{eqnarray}
where $r_i$ denotes the rank of $\rho_i$. Note for completeness that it has been already shown in \cite{raynal05a} that the rank of $E_?$ is upper bounded by the ranks of the two density matrices $\rho_0$ and $\rho_1$:
\begin{eqnarray}\label{r?}
r_{E_?} \le min(r_0,r_1).
\end{eqnarray}

We are now ready to derive our first theorem. For this theorem, we assume two density matrices without overlapping supports (this can always be achieved using the reduction mechanism of \cite{raynal03a}). The theorem states that the POVM elements $E_0$ and $E_1$ of an optimal USD measurement have rank $r_0$ and $r_1$, respectively, only if the two operators $\rho_0-\sqrt{\frac{\eta_1}{\eta_0}} \sqrt{\sqrt{\rho_0}\rho_1\sqrt{\rho_0}}$ and $\rho_1-\sqrt{\frac{\eta_0}{\eta_1}}\sqrt{\sqrt{\rho_1}\rho_0\sqrt{\rho_1}}$ are positive semi-definite. Here comes the precise statement.

\begin{theorem}\label{rank_theo}Constraints on the rank of the two POVM elements $E_0$ and $E_1$ of an optimal USD measurement\\
Consider a USD problem defined by two density matrices $\rho_0$ and $\rho_1$ and their respective {\rm a priori} probabilities $\eta_0$ and $\eta_1$ such that their supports satisfy ${\mathcal S}_{\rho_0} \cap {\mathcal S}_{\rho_1}=\{0\}$ (Any USD problem of two density matrices can be reduced to such a form according to \cite{raynal03a}). Consider also an optimal measurement $\{E_0^{opt},E_1^{opt}, E_?^{opt}\}$ to that problem. Let $F_0$ and $F_1$ be the two operators $\sqrt{\sqrt{\rho_0}\rho_1\sqrt{\rho_0}}$ and $\sqrt{\sqrt{\rho_1}\rho_0\sqrt{\rho_1}}$. The fidelity $F$ of the two states $\rho_0$ and $\rho_1$ is then given by $F=\Tr(F_0)=\Tr(F_1)$. Let $r_0$ and $r_1$ be the rank of the two density matrices $\rho_0$ and $\rho_1$.\\

If the POVM elements $E_0^{opt}$ and $E_1^{opt}$ have the same rank as the density matrices $\rho_0$ and $\rho_1$, respectively, then
\begin{eqnarray}
\left\{
\begin{array}{c}
\rho_0-\sqrt{\frac{\eta_1}{\eta_0}} F_0 \geq 0, \\
\rho_1-\sqrt{\frac{\eta_0}{\eta_1}} F_1 \geq 0.
\end{array}
\right.
\end{eqnarray}
\end{theorem}

The proof of this theorem relies on Eldar's necessary and sufficient conditions \cite{eldar04a} and tools developed in \cite{raynal05a}. A detailed proof is given in Appendix A. Interestingly Theorem~\ref{rank_theo} suggests that the two POVM elements $E_0$ and $E_1$ have rank $r_0$ and $r_1$, respectively, only in a small regime of the ratio $\sqrt{\frac{\eta_1}{\eta_0}}$ around $1$. Indeed the positivity of the two operators $\rho_0-\sqrt{\frac{\eta_1}{\eta_0}} F_0$ and $\rho_1-\sqrt{\frac{\eta_0}{\eta_1}}F_1$ is only possible when \cite{herzog05a,raynal05a}
\begin{eqnarray}
 \frac{\Tr(P_1 \rho_0)}{F}\le \sqrt{\frac{\eta_1}{\eta_0}} \le \frac{F}{\Tr(P_0 \rho_1)}.
\end{eqnarray}
Note that these boundaries can be made tighter if more knowledge on $\rho_0$ and $\rho_1$ is provided. Such an example of tighter bounds is given in Appendix B. Other boundaries were also derived in \cite{guo06a}.\\

Theorem~\ref{rank_theo} can also be rephrased as follows:\\

Whenever the two operators $\rho_0-\sqrt{\frac{\eta_1}{\eta_0}} \sqrt{\sqrt{\rho_0}\rho_1\sqrt{\rho_0}}$ and $\rho_1-\sqrt{\frac{\eta_0}{\eta_1}}\sqrt{\sqrt{\rho_1}\rho_0\sqrt{\rho_1}}$ are not positive semi-definite, at least one of the two POVM elements $E_i$, $i=0,1$, of an optimal USD measurement does not have rank $r_i$. In the case of a standard form \footnote{In the case of a standard form, we not only have ${\mathcal S}_{\rho_0} \cap {\mathcal S}_{\rho_1}=\{0\}$ but also ${\mathcal K}_{\rho_0} \cap {\mathcal S}_{\rho_1}=\{0\}$ and ${\mathcal K}_{\rho_1} \cap {\mathcal S}_{\rho_0}=\{0\}$.}, we can then show that:\\

If the two operators $\rho_0-\sqrt{\frac{\eta_1}{\eta_0}} \sqrt{\sqrt{\rho_0}\rho_1\sqrt{\rho_0}}$ and $\rho_1-\sqrt{\frac{\eta_0}{\eta_1}}\sqrt{\sqrt{\rho_1}\rho_0\sqrt{\rho_1}}$ are not positive semi-definite, then a two-dimensional subspace can be split off from the original USD problem.\\

This corollary of Theorem~\ref{rank_theo} actually is a fourth but incomplete reduction theorem. 'Reduction theorem' because no optimization is required onto that two-dimensional subspace. 'Incomplete', because the existence and the structure of this subspace are known but no complete analytical characterization is available yet. The precise statement of this Corollary follows.

\begin{corollary}\label{corollary}A fourth, incomplete, reduction theorem\\
Consider a standard USD problem defined by two density matrices $\rho_0$ and $\rho_1$ and their respective {\it a priori} probabilities $\eta_0$ and $\eta_1$ (any USD problem of two density matrices can be reduced to such a form according to \cite{raynal03a}). Consider also an optimal measurement $\{E_0^{opt},E_1^{opt}, E_?^{opt}\}$ to that problem. Let $F_0$ and $F_1$ be the two operators $\sqrt{\sqrt{\rho_0}\rho_1\sqrt{\rho_0}}$ and $\sqrt{\sqrt{\rho_1}\rho_0\sqrt{\rho_1}}$. The fidelity $F$ of the two states $\rho_0$ and $\rho_1$ is then given by $F=\Tr(F_0)=\Tr(F_1)$.\\

\begin{eqnarray}
\textrm{If one of the conditions}\,\,\,
\left\{ \begin{array}{c}
\rho_0-\sqrt{\frac{\eta_1}{\eta_0}} F_0 \geq 0 \\
\rho_1-\sqrt{\frac{\eta_0}{\eta_1}} F_1 \geq 0
\end{array}
\right. \textrm{is violated, then}
\end{eqnarray}
there exists a two-dimensional subspace that can be split off from the original Hilbert space.\\

This two-dimensional subspace is characterized by a two-dimensional orthonormal basis $\{|e\rangle,|e'\rangle\}$ such that either
\begin{eqnarray*}
|e\rangle \in {\cal S}_{\rho_0}\,\,\textrm{and}\,\, |e'\rangle \in {\cal K}_{\rho_0} \,\,\textrm{and}\,\,
\left\{
\begin{array}{c}
E_?^{opt} |e\rangle =|e\rangle\\
E_1^{opt} |e'\rangle=|e'\rangle\\
E_0^{opt} |e\rangle =E_0^{opt} |e'\rangle=E_1^{opt}|e\rangle=E_?^{opt} |e'\rangle=0,
\end{array}
\right.
\end{eqnarray*}
\begin{eqnarray*}
\textrm{or} \nonumber
\end{eqnarray*}
\begin{eqnarray*}
|e\rangle \in {\cal S}_{\rho_1}\,\,\textrm{and}\,\, |e'\rangle \in {\cal K}_{\rho_1} \,\,\textrm{and}\,\,
\left\{
\begin{array}{c}
E_?^{opt} |e\rangle =|e\rangle\\
E_0^{opt} |e'\rangle=|e'\rangle\\
E_1^{opt} |e\rangle =E_1^{opt} |e'\rangle=E_0^{opt}|e\rangle=E_?^{opt} |e'\rangle=0.
\end{array}
\right.
\end{eqnarray*}
\end{corollary}
First let us note that this corollary makes the assumption of a {\it standard} USD problem. It is in fact not necessary to make such a strong assumption to derive the existence of some eigenvector of $E_?$ and $E_{0/1}$ with eigenvalue $1$ because Theorem~\ref{rank_theo} is valid for any pair of density matrices without overlapping supports. Nevertheless Corollary~\ref{corollary} aims to be a fourth reduction theorem. It means in particular that, for any given USD problem of two density matrices, we would like to apply the four reduction theorems and always end up with the optimal USD measurement.\\

The above corollary is a kind of incomplete {\it reduction theorem}. A reduction theorem is a theorem that allows us to decrease the size of a USD problem by splitting off some subspace onto which no optimization is needed. To have a complete reduction theorem here, we would need to characterize $|e\rangle$ and $|e'\rangle$ without solving the whole optimization problem. So far the existence of $|e\rangle$ and $|e'\rangle$ is ensured and their structure is known (they are eigenvectors of $E_?$ and $E_{0/1}$ with eigenvalue 1). If $|e\rangle$ and $|e'\rangle$ were completely characterized in terms of $\rho_0$, $\rho_1$, $\eta_0$ and $\eta_1$, we would have a recipe to solve any USD problem. To see that, let us assume that $|e\rangle$ and $|e'\rangle$ can be fully characterized and start with two generic mixed states. In the following, the exponent $(r)$ denotes the rank of the density matrices after reduction.  First, we use the first three reduction theorems to bring the problem into its standard form. We then check whether the two operators $\rho_0^{(r)}-\sqrt{\frac{\eta_1^{(r)}}{\eta_0^{(r)}}} F_0^{(r)}$ and $\rho_1^{(r)}-\sqrt{\frac{\eta_0^{(r)}}{\eta_1^{(r)}}} F_1^{(r)}$ are positive semi-definite. If this is the case, then we know the optimal failure probability as well as the optimal measurement to perform since this case falls into the first class of exact solutions \cite{raynal05a}. If at least one of the two operators $\rho_0^{(r)}-\sqrt{\frac{\eta_1^{(r)}}{\eta_0^{(r)}}} F_0^{(r)}$ and $\rho_1^{(r)}-\sqrt{\frac{\eta_0^{(r)}}{\eta_1^{(r)}}} F_1^{(r)}$ is not positive semi-definite, we can use our last reduction theorem to get rid of a two-dimensional subspace and define the new density matrices $\rho_0^{(r-1)}$ and $\rho_1^{(r-1)}$ together with their respective {\it a priori} probabilities $\eta_0^{(r-1)}$ and $\eta_1^{(r-1)}$. At that point we check again the positivity of the two operators $\rho_0^{(r-1)}-\sqrt{\frac{\eta_1^{(r-1)}}{\eta_0^{(r-1)}}} F_0^{(r-1)}$ and $\rho_1^{(r-1)}-\sqrt{\frac{\eta_0^{(r-1)}}{\eta_1^{(r-1)}}} F_1^{(r-1)}$ of the reduced problem. We see here a constructive way to solve any USD problem. If the two operators $\rho_0^{(r')}-\sqrt{\frac{\eta_1^{(r')}}{\eta_0^{(r')}}} F_0^{(r')}$ and $\rho_1^{(r')}-\sqrt{\frac{\eta_0^{(r')}}{\eta_1^{(r')}}} F_1^{(r')}$ never happen to be positive, we end up with only two pure states and can finally find the optimal measurement (see Fig.~\ref{4red}). A detailled proof of Corollary~\ref{corollary} is given in Appendix C.\\

\begin{figure}[h!]
  \centering
  \includegraphics[width=10cm]{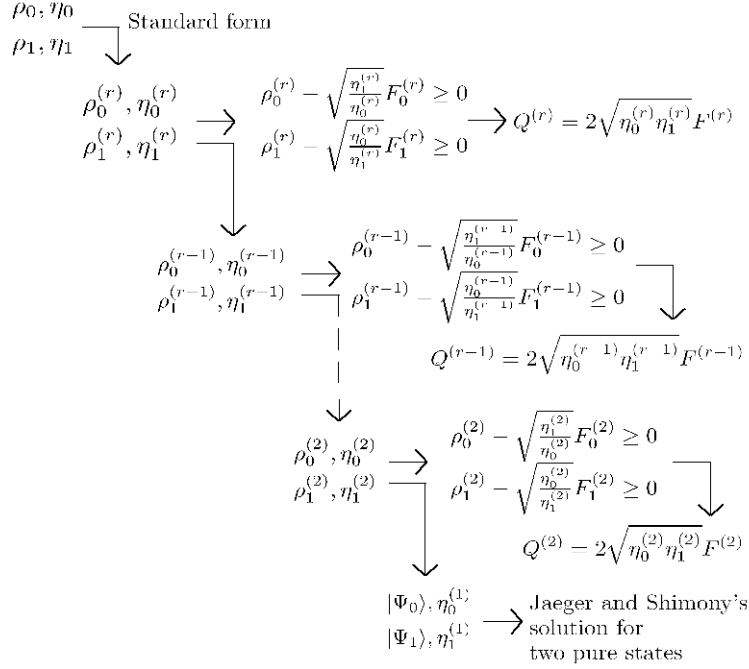}
  \caption{A constructive way to solve any USD problem if $|e\rangle$ and $|e'\rangle$ in Corollary 1 were fully characterized (the exponent $^{(r)}$ denotes the rank of the density matrices after reduction)}
  \label{4red}
\end{figure}

So far, there are only two ways to find a complete characterization of the two eigenvectors $|e\rangle$ and $|e'\rangle$ involved in Corollary~{\ref{corollary}}. The first possibility is to consider a low-dimensional USD problem. The second option is to consider a highly symmetric problem. The former case simply is the case of the two pure states where we want to unambiguously discriminate two pure states $|\Psi_0\rangle$ and $|\Psi_1\rangle$ with probabilities $\eta_0$ and $\eta_1$. Either the operators $\rho_0-\sqrt{\frac{\eta_1}{\eta_0}} \sqrt{\sqrt{\rho_0}\rho_1\sqrt{\rho_0}}$ and $\rho_1-\sqrt{\frac{\eta_0}{\eta_1}} \sqrt{\sqrt{\rho_1}\rho_0\sqrt{\rho_1}}$ are positive semi-definite or we have $|e\rangle \in {{\cal S}_{\rho_{0/1}}}$ eigenvector of $E_?$ with eigenvalue 1, and $|e'\rangle \in {{\cal K}_{\rho_{0/1}}}$ eigenvector of $E_{1/0}$ with eigenvalue 1. In only two dimensions, there is no freedom and $|e\rangle$ and $|e'\rangle$ must be $|\Psi_{0/1}\rangle$ and $|\Psi_{0/1}^\perp\rangle$. The two pure states case is solved extremely elegantly and rapidly thanks to Corollary~\ref{corollary}. If we are interested in higher dimensions we must consider a symmetry to give us enough constraints to fully characterize $|e\rangle$ and $|e'\rangle$. With the help of the Geometrically Uniform (GU) symmetry it is possible to go up to four dimensions and obtain a second class of exact solutions. This solution of generic USD problem is the object of the next section.

\section{Second class of exact solutions}\label{class}

Before deriving the main result of this paper, we need to introduce the so-called {\it Geometrically Uniform} (GU) states, often met in practical applications \footnote{In a cryptographic context, the {\it bit value} states and {\it basis} states in the BB84-type protocol using weak coherent pulses and a phase reference exhibit such a GU symmetry.}. While symmetric states are generated from one generator state and a single unitary transformation, GU states are generated from one generator and a group of unitaries \cite{eldar01a,eldar02a,eldar03a,eldar03b,eldar03c,eldar04a}. More precisely, a set of GU states is a set of density matrices $\{ \rho_i\}$, $i=0,...,n-1$ such that $\rho_i=U_i \rho U_i^\dagger$ where $\rho$ is an arbitrary density matrix called the generator and the set $\{U_i\}$, $i=0,...,n-1$ is a set of unitary matrices that form an Abelian group. In order not to break the symmetry of the states, we assume that the {\it a priori} probabilities are all equal to $\frac{1}{n}$. A consequence of the group structure of the set $\{U_i\}$ is that we can always consider $U_0$ as the identity and $\rho_0$ as the generator for a given set of GU states. We can therefore always write two GU states as $\rho_0$ and $\rho_1 = U \rho_0 U$, where $U$ is an involution (i.e.\ a unitary transformation $U$ such that $U^2={\openone}$ and $U^\dagger=U$) with $\eta_0=\eta_1=\frac{1}{2}$. Let us finally note that two GU states are two symmetric states since only a single unitary is needed.\\

We can now present the main result of this paper, that is, the optimal failure probability for unambiguously discriminating two {\it geometrically uniform} states in a four dimensional Hilbert space and its corresponding optimal measurement.

\begin{theorem}\label{class_theo}Optimal unambiguous discrimination of two geometrically uniform states in four dimensions\\
Consider a USD problem defined by two {\it geometrically uniform} states $\rho_0$ and $\rho_1$ of rank 2 with equal {\it a priori} probabilities and spanning a four-dimensional Hilbert space. Let $F_0$ and $F_1$ be the two operators $\sqrt{\sqrt{\rho_0}\rho_1\sqrt{\rho_0}}$ and $\sqrt{\sqrt{\rho_1}\rho_0\sqrt{\rho_1}}$. The fidelity $F$ of the two states $\rho_0$ and $\rho_1$ is then given by $F=\Tr(F_0)=\Tr(F_1)$. We denote by $P_0^\perp$ and $P_1^\perp$ the projectors onto the kernel of $\rho_0$ and $\rho_1$. The optimal failure probability $Q^{\textrm{opt}}$ for USD then satisfies 

\begin{eqnarray}
Q^{\mathrm{opt}}=
\left\{ \begin{array}{l}
F \\
1-\langle x|\rho_0|x\rangle
\end{array}\right.
\left. \begin{array}{l}
\textrm{if}\,\,\,\, \rho_0-F_0 \geq 0 \\
\textrm{otherwise,}
\end{array}\right.
\end{eqnarray}

where $P^{\perp}_{1}\,U\,P^{\perp}_{1}=a|0\rangle\langle0|-b|1\rangle\langle1|$ and $|x \rangle=\frac{1}{\sqrt{a+b}}(e^{iArg(\langle 0 | \rho_0| 1 \rangle)}\sqrt{b}|0\rangle + \sqrt{a} |1\rangle)$.\\
The POVM elements that realize these optimal failure probabilities are given in the two respective cases by\\
\begin{eqnarray}
1.\,\,\,\, E_0&=&\Sigma^{-1} \sqrt{\rho_0} \left(\rho_0- F_0 \right) \sqrt{\rho_0}\Sigma^{-1}\,\,\, \textrm{where} \,\,\, \Sigma=\rho_0+\rho_1\\ \nonumber
E_1&=&U E_0 U \\ \nonumber
E_?&=&{\openone}- E_0 - U E_0 U\\ \nonumber
\\ \nonumber
\\ \nonumber
2.\,\,\,\, E_0&=&|x \rangle \langle x |\\ \nonumber
E_1&=&U E_0 U \\ \nonumber
E_?&=&{\openone}- E_0 - U E_0 U
\end{eqnarray}
\end{theorem}

\paragraph*{\bf Proof}
We consider a USD problem defined by two {\it geometrically uniform} states $\rho_0$ and $\rho_1=U \rho_0 U$, $U^2={\openone}$, of rank 2, spanning a four-dimensional Hilbert space. Since $rank(\rho_0)+rank(\rho_1)=rank(\rho_0+\rho_1)$, the two supports do not overlap \cite{raynal06a}. Note that this problem is in the standard form as soon as it has no trivial orthogonal directions and no block diagonal structures.\\

The symmetry between the two states $\rho_0$ and $\rho_1$ allows further simplifications.  Actually, Eldar proved in \cite{eldar04a} that the optimal measurement to unambiguously discriminate {\it geometrically uniform} states can be chosen {\it geometrically uniform}, too. Thus the POVM elements have the form:
\begin{eqnarray}
&E_0&,\\ \nonumber
&E_1&=U E_0 U,\\ \nonumber
&E_?&={\openone}- E_0 - U E_0 U.
\end{eqnarray}
In addition we know from the first class of exact solutions \cite{raynal05a} that, for two density matrices $\rho_0$ and $\rho_1$ with equal {\it a priori} probabilities and without overlapping supports,
\begin{eqnarray}
Q^{\mathrm{opt}} = F  & \Leftrightarrow & \, 
\left\{\begin{array}{cc}
\rho_0-F_0 \ge 0 \\ 
\rho_1-F_1 \ge 0 \\ 
\end{array}\right. .
\end{eqnarray}
This is even a stronger statement than the desired one since we have an equivalence where we only want an implication \footnote{The implication from the right to the left is the only important direction for our purpose. The assumption ${\mathcal S}_{\rho_0} \cap {\mathcal S}_{\rho_1}=\{0\}$ is required to prove that if $\rho_0-F_0 \ge 0$ then $Q^{\mathrm{opt}} = F$. Without this assumption, only the other direction is true (See \cite{raynal05a} for more details).}. Due to the symmetry of the states, the operators $\rho_0-F_0$ and $\rho_1-F_1$ share the same spectrum and the above conditions reduce to
\begin{eqnarray}
Q^{\mathrm{opt}} = F  & \Leftrightarrow & \, 
\begin{array}{cc}
\rho_0-F_0 \ge 0. \\ 
\end{array}
\end{eqnarray}
If $\rho_0-F_0 \ngeq 0$, then Theorem~\ref{rank_theo} tells us that at least one of the two POVM elements $E_0$ and $E_1$ does not have rank 2, the rank of the two density matrices $\rho_0$ and $\rho_1$. Since $E_1=U E_0 U$, $E_0$ and $E_1$ have the same rank so that if $\rho_0-F_0 \ngeq 0$ then $r_{E_0}=r_{E_1} < 2$. If $r_{E_0}=r_{E_1}=0$ then $E_?= {\openone}$ and $Q=1$, which is clearly not the optimal solution as the density matrices are linear independent. Let us now focus on the remaining case $r_{E_0}=r_{E_1}=1$.\\

First of all it is not too difficult to see that an optimal USD measurement such that $r_{E_0}=r_{E_1}=1$ and $rank(E_?) \le 2$ must be a projective measurement with $rank(E_?)=2$ (See Appendix D for details). For such measurements, we have $\Tr(E_0 E_1)=0$ or simply $\langle x |U| x \rangle=0$ where $E_0=| x \rangle \langle x |$. Since $| x \rangle$ lies in ${\mathcal K}_{\rho_1}$, this last relation is equivalent to
\begin{eqnarray}\label{spectop}
\langle x |P_1^\perp U P_1^\perp| x \rangle=0.
\end{eqnarray}
Importantly, the operator $P_1^\perp U P_1^\perp$ has one positive and one negative eigenvalue whenever $\rho_0-F_0$ is not positive semi-definite (a detailed proof can be found in Appendix E). In the eigenbasis $\{| 0 \rangle,| 1 \rangle\}$ of $P_1^\perp U P_1^\perp$, we write 
\begin{eqnarray}
| x \rangle=
\left(\begin{array}{c}
\alpha\\
\beta
\end{array}
\right)
\end{eqnarray}
and
\begin{eqnarray}
P_1^\perp U P_1^\perp=
\left(\begin{array}{cc}
a&0\\
0&-b
\end{array}
\right),\,\,\, a,b>0.
\end{eqnarray}
If we also consider the normalization of $| x \rangle$, we end up with the following system of two equations:
\begin{eqnarray}
\left\{\begin{array}{c}
|\alpha|^2 a - |\beta|^2 b=0\\
|\alpha|^2 + |\beta|^2 =1
\end{array}
\right. .
\end{eqnarray}
Up to a global phase, it admits a family of solutions parametrized by a relative phase $\Phi \in [0;2\pi[$:
\begin{eqnarray}
\{\alpha=\frac{e^{i\Phi}}{\sqrt{1+a/b}},\beta=\frac{1}{\sqrt{1+b/a}}\}.
\end{eqnarray}
In the eigenbasis $\{| 0 \rangle,| 1 \rangle\}$ we can therefore write
\begin{eqnarray}
| x \rangle=
\left(\begin{array}{c}
\frac{e^{i\Phi}}{\sqrt{1+a/b}}\\
\frac{1}{\sqrt{1+b/a}}
\end{array}
\right).
\end{eqnarray}
We then use again the fact that we are interested in  the optimal measurement. Note that we already invoked the optimality condition when we used Theorem~\ref{rank_theo} to state that $r_{E_0}=r_{E_1}<2$ whenever $\rho_0-F_0 \ngeq 0$. So far $|x \rangle$ is a valid choice for any USD projective measurement that is GU symmetric. To find the optimal one, we evaluate the corresponding success probability $P_{success}$. Because of the symmetry of the two GU states, $\Tr(E_0 \rho_0)=\Tr(E_1 \rho_1)$ and  the success probability for unambiguously discriminating the two GU state $\rho_0$ and $\rho_1$ takes the form
\begin{eqnarray}
P_{success}=\Tr(E_0 \rho_0)=\langle x |\rho_0| x \rangle.
\end{eqnarray}
After a short calculation, we obtain
\begin{eqnarray}
P_{success}= \frac{1}{a+b}\left(b\langle0|\rho_0|0\rangle+a\langle1|\rho_0|1\rangle+2\sqrt{ab}Re(\langle0|\rho_0|1\rangle e^{-i\Phi})\right).
\end{eqnarray}
To maximize this success probability, we choose $\Phi$ such that $Re(\langle0|\rho_0|1\rangle e^{-i\Phi})=|\langle0|\rho_0|1\rangle|$. $\Phi$ must therefore be equal to $Arg(\langle0|\rho_0|1\rangle)$ and finally
\begin{eqnarray}
| x \rangle=
\left(\begin{array}{c}
\frac{e^{iArg(\langle0|\rho_0|1\rangle)}}{\sqrt{1+a/b}}\\
\frac{1}{\sqrt{1+b/a}}
\end{array}
\right).
\end{eqnarray} This completes the proof. \hfill $\blacksquare$\\

Let us remark here that as soon as $\rho_0-F_0 \ngeq 0$ the optimal measurement is $2$x$2$ block diagonal even if the two states $\rho_0$ and $\rho_1$ are not. In the next section we consider an example of both theoretical and practical interest. In fact, we consider the {\it Bennett and Brassard 1984} protocol (BB84 protocol) implemented through weak coherent pulses with a strong phase reference.

\section{Application of the second class of exact solutions to the BB84 protocol}\label{BB84}

The Bennett and Brassard 1984 cryptographic protocol \cite{bennett84a} provides a method to distribute a private key between two parties and therefore allow an unconditionally secure communication. We consider in this section the implementation of a BB84-type Quantum Key Distribution (QKD) protocol that uses weak coherent pulses with a phase reference \cite{dusek00a}. In that context, two important questions related to unambiguous state discrimination can be addressed. First, 'With what probability can an eavesdropper unambiguously distinguish the {\it basis} of the signal?' and second 'With what probability can an eavesdropper unambiguously determine which {\it bit value} is sent without being interested in the knowledge of the basis?' These two questions can be translated in some unambiguous discrimination task concerning two {\it geometrically uniform} mixed states in a four-dimensional Hilbert space. We answer these two questions providing useful insights for further investigations on practical implementations of Quantum Key Distribution protocols. Note that the details of all the following calculations can be found in \cite{raynal06a}.\\

\subsection{The pairs of mixed states to discriminate}

The implementation of the BB84 considered here makes use of the four quantum optical coherent states $\{|\pm~\alpha~\rangle , |\pm~i~\alpha~\rangle \}$ \cite{dusek00a}, where a coherent state $| \beta \rangle$ is characterized by its complex amplitude  $\beta \in \mathbb{C}$. The first question mentioned above refers to the unambiguous discrimination of the two {\it basis} mixed states
\begin{eqnarray}
\rho_r&=&\frac{1}{2}\left(|\alpha \rangle \langle \alpha|+|-\alpha \rangle \langle -\alpha|\right),\\
\rho_i&=&\frac{1}{2}\left(| i \alpha \rangle \langle i \alpha|+| -i \alpha \rangle \langle -i \alpha|\right).
\end{eqnarray}
The second question refers to the unambiguous discrimination of the two {\it bit value} mixed states
\begin{eqnarray}
\rho_0&=&\frac{1}{2}\left(|\alpha \rangle \langle \alpha|+|i\alpha \rangle \langle i\alpha|\right),\\
\rho_1&=&\frac{1}{2}\left(|-\alpha \rangle \langle -\alpha|+|-i\alpha \rangle \langle -i\alpha|\right).
\end{eqnarray}

Following the idea presented in \cite{dusek00a}, we can write these four density matrices in a four-dimensional Hilbert space
\begin{eqnarray}
\rho_r=\left(
\begin{array}{cccc}
c_0^2 & 0 & c_0c_2 & 0\\
0 & c_1^2 & 0 & c_1c_3\\
c_0c_2 & 0 & c_2^2 & 0\\
0 & c_1c_3 & 0 & c_3^2 \end{array}
\right) , \end{eqnarray}
\begin{eqnarray}
\rho_i=\left(
\begin{array}{cccc}
c_0^2 & 0 & -c_0c_2 & 0\\
0 & c_1^2 & 0 & -c_1c_3\\
-c_0c_2 & 0 & c_2^2 & 0\\
0 & -c_1c_3 & 0 & c_3^2 \end{array}\right),
\end{eqnarray}
\begin{eqnarray}
\rho_0=\left(
\begin{array}{cccc}
c_0^2 & \frac{1-i}{2} c_0c_1 & 0 & \frac{1+i}{2} c_0c_3\\
\frac{1+i}{2} c_1c_0 & c_1^2 & \frac{1-i}{2} c_1c_2 & 0\\
0 & \frac{1+i}{2} c_2c_1 & c_2^2 & \frac{1-i}{2} c_2c_3\\
\frac{1-i}{2} c_3c_0 & 0 & \frac{1+i}{2} c_3c_2 & c_3^2 \end{array}\right),
\end{eqnarray}
and
\begin{eqnarray}
\rho_1=\left(
\begin{array}{cccc}
c_0^2 & -\frac{1-i}{2} c_0c_1 & 0 & -\frac{1+i}{2} c_0c_3\\
-\frac{1+i}{2} c_1c_0 & c_1^2 & -\frac{1-i}{2} c_1c_2 & 0\\
0 & -\frac{1+i}{2} c_2c_1 & c_2^2 & -\frac{1-i}{2} c_2c_3\\
-\frac{1-i}{2} c_3c_0 & 0 & -\frac{1+i}{2} c_3c_2 & c_3^2 \end{array}\right)
\end{eqnarray}
where the coefficients $c_i$'s can be expressed as functions of the mean photon number $\mu=|\alpha|^2$ of the coherent pulse \cite{dusek00a}:
\begin{eqnarray}
c_0=\frac{1}{\sqrt{2}}e^{-\frac{\mu}{2}}\sqrt{\cosh(\mu)+\cos(\mu)},\\
c_1=\frac{1}{\sqrt{2}}e^{-\frac{\mu}{2}}\sqrt{\sinh(\mu)+\sin(\mu)},\\
c_2=\frac{1}{\sqrt{2}}e^{-\frac{\mu}{2}}\sqrt{\cosh(\mu)-\cos(\mu)},\\
c_3=\frac{1}{\sqrt{2}}e^{-\frac{\mu}{2}}\sqrt{\sinh(\mu)-\sin(\mu)}.
\end{eqnarray}

\subsection{USD of the {\it basis} mixed states}
The first question refers to the USD of the two density matrices $\rho_r$ and $\rho_i$. One can actually calculate the spectrum of the operator $\rho_r-F_r$ and find that
\begin{eqnarray}
Spect(\rho_r-F_r)=\{max\{c_0^2,c_2^2\},max\{c_1^2,c_3^2\}\}\}
\end{eqnarray}
which is positive for any value of $ \mu$. Therefore Theorem~\ref{class_theo} tells us that $Q^{opt}=F$ for any value of $ \mu$. One can also calculate the fidelity which takes the form $F=|c_0^2 - c_2^2| +|c_1^2 - c_3^2|$. In terms of the mean photon number the optimal failure probability (see Fig.~\ref{bb84q1}) is finally expressed as
\begin{eqnarray}
Q^{opt}=e^{-\mu}\left(|\cos \mu |+|\sin \mu |\right), \forall \mu.
\end{eqnarray}

\begin{figure}[h!]
  \centering
  \includegraphics[width=7cm]{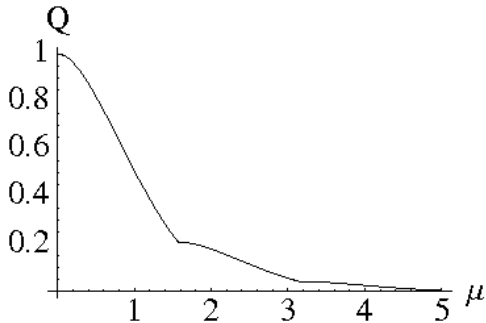}
  \caption{Optimal failure probability for USD of the {\it basis} mixed states ($\mu$ in photons per pulse)}
  \label{bb84q1}
\end{figure}

\subsection{USD of the {\it bit value} mixed states}
The second question refers to the USD of the two density matrices $\rho_0$ and $\rho_1$. The spectrum of the operator $\rho_0-F_0$ is now given by
\begin{eqnarray}
Spect(\rho_0-F_0)=\frac{1}{2}\left(1-e^{-\mu} \pm e^{-2\mu}\sqrt{1+e^{2\mu}-2e^{\mu}\cos(2\mu)}\right).
\end{eqnarray}
This spectrum is not always positive (see Fig.~\ref{spectrum1}).
\begin{figure}[h!]
  \centering
  \includegraphics[width=7cm]{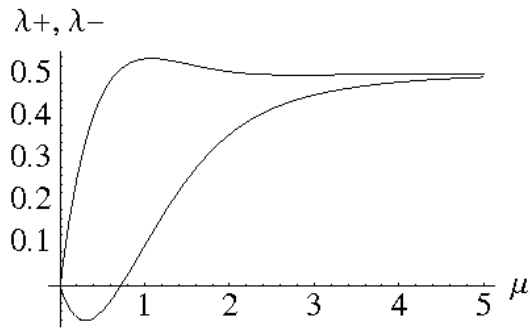}
  \caption{Spectrum of the operator $\rho_0-F_0$ for USD of the {\it bit value} mixed states ($\mu$ in photons per pulse)}
  \label{spectrum1}
\end{figure}
Indeed only in a regime of relatively large $\mu$ ($\mu \ge \mu_0\thickapprox0.7193$ photon per pulse}), the quantity $\frac{1}{2}(1-e^{-\mu} - e^{-2\mu}\sqrt{1+e^{2\mu}-2e^{\mu}\cos(2\mu)})$ is greater than $0$.
In the regime $\mu \ge \mu_0$, the positivity of the operator $\rho_0-F_0$ ensures that the optimal failure probability reaches the fidelity bound $F$, which can be expresses as $e^{-\mu}$ (See \cite{raynal06a} for details). We therefore obtain 
\begin{eqnarray}
Q^{\mathrm{opt}}=e^{-\mu}, \forall \mu \ge \mu_0.
\end{eqnarray}


Note that for $\mu=\mu_0$, the POVM elements $E_0$ and $E_1$ have rank $1$ since one eigenvalue of $\rho_0-F_0$ vanishes.\\

\begin{figure}[h!]
  \centering
  \includegraphics[width=7cm]{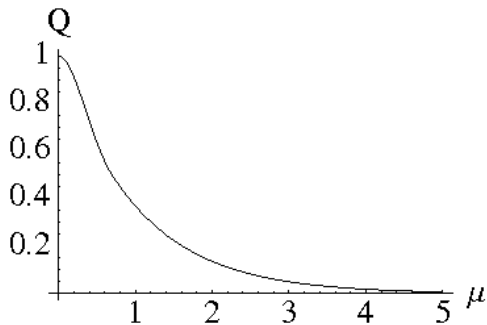}
  \caption{Optimal failure probability for USD of the {\it bit value} mixed states ($\mu$ in photons per pulse)}
  \label{bb84q4}
\end{figure}

In the regime $\mu < \mu_0$ where the operator $\rho_0-F_0$ is not positive semi-definite Theorem~\ref{class_theo} tells us that the optimal failure probability for unambiguously discriminating the {\it bit value} mixed states is
\begin{eqnarray}
Q^{\mathrm{opt}} = 1-\frac{1}{a+b}(b\langle0|\rho_0|0\rangle+a\langle1|\rho_0|1\rangle+2\sqrt{ab}|\langle0|\rho_0|1\rangle|)
\end{eqnarray}
where $P^{\perp}_{1}\,U\,P^{\perp}_{1}=a|0\rangle\langle0|-b|1\rangle\langle1|$ and $|x \rangle=\frac{1}{\sqrt{a+b}}(e^{iArg(\langle 0 | \rho_0| 1 \rangle)}\sqrt{b}|0\rangle + \sqrt{a} |1\rangle)$. The complete graph for the failure probability is shown in Fig.~\ref{bb84q4}.\\

\begin{figure}[h!]
  \centering
  \includegraphics[width=7cm]{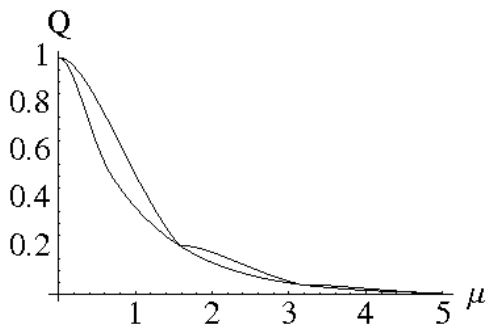}
  \caption{Comparison between the optimal failure probabilities for USD of the {\it basis} and the {\it bit value} mixed states ($\mu$ in photons per pulse)}
  \label{bb84qccl}
\end{figure}

So far no neat expression for the optimal failure probability is known  in terms of $\mu$  for $\mu < \mu_0$. This comes from the rather complicated form of the states $\rho_0$ and $\rho_1$ and the fact that no analytical expression of $P_0$ and $P_1$ is known. In Fig.~\ref{bb84qccl} we finally compare the optimal failure probabilities for USD of the {\it basis} and the {\it bit value} mixed states.

\section{Conclusion}\label{conclusion}

In this paper we have provided a theorem giving constraints on the rank of the conclusive outcomes of an optimal USD measurement. This theorem also implies a partial fourth reduction theorem. This corollary tells us that either the failure probability equals the overall lower bound $2\sqrt{\eta_0 \eta_1}F$ or a two-dimensional subspace can be split off from the original Hilbert space. In fact, this result can be used as a toolbox to go beyond special cases and solve analytically USD problems that can not be solved with the help of the three reduction theorems. We have then employed these results to derive a second class of exact solutions. This class corresponds to any pair of equally probable geometrically uniform states in a four-dimensional Hilbert space. For that class of states we give the optimal failure probability as well as the associated optimal measurement. As an application, we have used this result to address two questions related to the implementation of the BB84 QKD protocol with weak coherent states. The questions 'With what probability can an eavesdropper unambiguously distinguish the {\it basis} of the signal?' and 'With what probability can an eavesdropper unambiguously determine which {\it bit value} is sent without being interested in the knowledge of the basis?' are each related to the unambiguous discrimination of a pair of geometrically uniform states in a four-dimensional Hilbert space.

\begin{center}
{\bf Acknowledgments}
\end{center}

We would like to thank Matthias Kleinmann, Hermann Kampermann and Dagmar Bruss for useful discussions. This work was supported by the DFG under the Emmy-Noether program and the EU Integrated Project QAP.\\

\cleardoublepage

\section{Appendices}

\subsection*{Appendix A: \bf Proof of Theorem~\ref{rank_theo}}
We now prove Theorem~\ref{rank_theo} which is concerned with an optimal measurement for unambiguously discriminating two mixed states $\rho_0$ and $\rho_1$. We can therefore use the necessary and sufficient conditions derived by Eldar in \cite{eldar04a}. We recall them here in a language adapted to our calculations.\\

A necessary and sufficient condition for a measurement $\{ E_k \}$, $k \in \{0,1,?\}$, to be optimal is the existence of a positive semi-definite operator $Z$ such that
\begin{eqnarray}\label{NSC}
Z E_?=0, \label{eq1}\\
E_0 (Z- \eta_0 \rho_0) E_0 =0, \label{eq2}\\
E_1 (Z- \eta_1 \rho_1) E_1 =0, \label{eq3}\\
P_1^\perp(Z- \eta_0 \rho_0)P_1^\perp \ge 0, \label{eq4}\\
P_0^\perp(Z- \eta_1 \rho_1)P_0^\perp \ge 0 \label{eq5},
\end{eqnarray}
where $P_i^\perp$ is the projector onto the kernel of $\rho_i$, $i=0,1$.\\

The proof of Theorem~\ref{rank_theo} can be decomposed in five steps. The first step corresponds to the restriction of Eldar's necessary and sufficient conditions to the case where the POVM elements $E_0$ and $E_1$ of a USD measurement have rank $r_0$ and $r_1$, respectively. After some calculations, this will provide us with a first statement\\

If the two POVM elements $E_0$ and $E_1$ of an optimal USD measurement have rank $r_0$ and $r_1$, respectively, then 
\begin{eqnarray}
\exists Z\ge 0 \,\,\,\textrm{such that}\,\,\,
\left\{\begin{array}{c}
ZE_?=0\\
P_0^\perp(Z-\eta_1 \rho_1)P_0^\perp=0\\
P_1^\perp(Z-\eta_0 \rho_0)P_1^\perp=0\\
\end{array}
\right. .
\end{eqnarray}
Therefore, to prove Theorem~\ref{rank_theo}, we will simplify have to show the equivalence:
\begin{eqnarray}
\exists Z\ge 0 \,\,\,\textrm{such that}\,\,\,
\left\{\begin{array}{c}
ZE_?=0\\
P_0^\perp(Z-\eta_1 \rho_1)P_0^\perp=0\\
P_1^\perp(Z-\eta_0 \rho_0)P_1^\perp=0\\
\end{array}
\right.
\Leftrightarrow
\left\{
\begin{array}{c}
\rho_0-\sqrt{\frac{\eta_1}{\eta_0}} F_0 \geq 0 \\
\rho_1-\sqrt{\frac{\eta_0}{\eta_1}} F_1 \geq 0
\end{array}
\right. .
\end{eqnarray}
The remaining four steps will show this equivalence. More precisely, the second step makes use of the notion of parallel addition. The third step uses the positive semi-definiteness of the unknown operator $Z$ and focuses on the equation $ZE_?=0$. The fourth step is concerned with the optimal failure probability. The fifth and final step uses an equivalence already shown in \cite{raynal05a}.\\

\noindent{\bf First step}\\
In that first step, we only want to prove an implication. We now see the following:\\

If the USD POVM is optimal, i.e.\ Eqn.(\ref{eq1})-(\ref{eq5}) are fulfilled, and $E_0$ and $E_1$ have rank $r_0$ and $r_1$, respectively, then the two Hermitian operators $P_1^\perp(Z-\eta_0 \rho_0)P_1^\perp$ and $P_0^\perp(Z-\eta_1 \rho_1)P_0^\perp$ must vanish. Indeed the situation is the following. We consider two positive semi-definite operators $A$ and $B$, with $A$ having full rank and $ABA^\dagger=0$. We can see this relation as the form $CC^\dagger=0$ with $C=A\sqrt{B}$. Such an equation $CC^\dagger=0$ is equivalent to $C=0$ for any matrix $C$. Consequently, $ABA^\dagger=0$ is equivalent to $A\sqrt{B}=0$. Finally, since $A$ is full rank, $A^{-1}$ exists and $B$ must vanish.\\

Let us now focus on Eqn.(\ref{eq2}) where $E_0$ and $P_1^\perp(Z-\eta_0 \rho_0)P_1^\perp$ both have support in ${\cal K}_{\rho_1}$. We assume $r_{E_0}=dim({\cal K}_{\rho_1})$($=r_0$ since the supports of $\rho_0$ and $\rho_1$ do not overlap). so that on the subspace ${\cal K}_{\rho_1}$ we can consider $E_0$ as being full rank. We then set $A=E_0$ and $B=P_1^\perp(Z-\eta_0 \rho_0)P_1^\perp$. Eqn.(\ref{eq2}) tells us that $ABA^\dagger=0$ with $A$ full rank on ${\cal K}_{\rho_1}$ thus $B=P_1^\perp(Z-\eta_0 \rho_0)P_1^\perp$ must vanish.\\
We can follow the same idea for Eqn.(\ref{eq3}).  We set $A=E_1$ and $B=P_0^\perp(Z-\eta_1 \rho_1)P_0^\perp$. Since we assume $r_{E_1}=dim({\cal K}_{\rho_0})$($=r_1$ since the supports of $\rho_0$ and $\rho_1$ do not overlap), $E_1$ can be considered as being full rank in ${\cal K}_{\rho_0}$ and therefore $P_0^\perp(Z-\eta_1 \rho_1)P_0^\perp$ must vanish.\\
Therefore we obtain that:\\

If the two POVM elements $E_0$ and $E_1$ of an optimal USD measurement have rank $r_0$ and $r_1$, respectively, then 
\begin{eqnarray}
\exists Z\ge 0 \,\,\,\textrm{such that}\,\,\,
\left\{\begin{array}{c}
ZE_?=0\\
P_0^\perp(Z-\eta_1 \rho_1)P_0^\perp=0 \label{qe1}\\
P_1^\perp(Z-\eta_0 \rho_0)P_1^\perp=0 \label{qe2}\\
P_0^\perp(Z- \eta_1 \rho_1)P_0^\perp \ge 0 \label{qe3}\\
P_1^\perp(Z- \eta_0 \rho_0)P_1^\perp \ge 0 \label{qe4}\\
\end{array}
\right. .
\end{eqnarray}

The above statement finally implies that:\\

If the two POVM elements $E_0$ and $E_1$ of an optimal USD measurement have rank $r_0$ and $r_1$, respectively, then 
\begin{eqnarray}
\exists Z\ge 0 \,\,\,\textrm{such that}\,\,\,
\left\{\begin{array}{c}
ZE_?=0\\
P_0^\perp(Z-\eta_1 \rho_1)P_0^\perp=0\\
P_1^\perp(Z-\eta_0 \rho_0)P_1^\perp=0\\
\end{array}
\right. .
\end{eqnarray}

It now remains to show the following equivalence to prove Theorem~\ref{rank_theo}:
\begin{eqnarray}\label{eqbob}
\exists Z\ge 0 \,\,\,\textrm{such that}\,\,\,
\left\{\begin{array}{c}
ZE_?=0\\
P_0^\perp(Z-\eta_1 \rho_1)P_0^\perp=0\\
P_1^\perp(Z-\eta_0 \rho_0)P_1^\perp=0\\
\end{array}
\right.
\Leftrightarrow
\left\{
\begin{array}{c}
\rho_0-\sqrt{\frac{\eta_1}{\eta_0}} F_0 \geq 0 \\
\rho_1-\sqrt{\frac{\eta_0}{\eta_1}} F_1 \geq 0
\end{array}
\right. .
\end{eqnarray}

\noindent{\bf Second step}\\
Since the supports of the two density matrices $\rho_0$ and $\rho_1$ do not overlap, we can make use of the notion of parallel addition and introduce the full rank operator $\Sigma^{-1}=(\rho_0 + \rho_1)^{-1}$ \cite{raynal05a}. Its main property lies in the relation
\begin{eqnarray}\label{sigma}
\rho_i \Sigma^{-1} \rho_j= \rho_i \delta_{ij},\,\,\,i=0,1.
\end{eqnarray}
As a consequence we get the interesting equalities
\begin{eqnarray}
\rho_0 \Sigma^{-1}&=&\rho_0 \Sigma^{-1} P_1^\perp, \label{eqn1}\\
P_1^\perp \rho_0 \Sigma^{-1}&=&P_1^\perp. \label{eqn2}
\end{eqnarray}
Indeed $\rho_0 \Sigma^{-1}=\rho_0 \Sigma^{-1}(P_1+P_1^\perp)=\rho_0 \Sigma^{-1}\rho_1 \rho_1^{-1} + \rho_0 \Sigma^{-1} P_1^\perp=\rho_0 \Sigma^{-1} P_1^\perp$. Moreover, $P_1^\perp=P_1^\perp {\openone}=P_1^\perp(\rho_0 +\rho_1) \Sigma^{-1}=P_1^\perp \rho_0 \Sigma^{-1}$.\\

If we now consider the equation $P_1^\perp(Z-\eta_0 \rho_0)P_1^\perp=0$, we can multiply it on the left-hand side by $\rho_0 \Sigma^{-1}$ and on the right-hand by $\Sigma^{-1} \rho_0$. We then have
\begin{eqnarray}
\rho_0 \Sigma^{-1} P_1^\perp(Z-\eta_0 \rho_0)P_1^\perp \Sigma^{-1} \rho_0=0.
\end{eqnarray}
Because of Eqn.~(\ref{eqn1}), this implies
\begin{eqnarray}
\rho_0 \Sigma^{-1}(Z-\eta_0 \rho_0) \Sigma^{-1} \rho_0=0.
\end{eqnarray}
We can also go in the other direction. Indeed it follows from the previous equation that:
\begin{eqnarray}
P_1^\perp \rho_0 \Sigma^{-1}(Z-\eta_0 \rho_0) \Sigma^{-1} \rho_0 P_1^\perp=0
\end{eqnarray}
which together with Eqn.~(\ref{eqn2}) yields
\begin{eqnarray}
P_1^\perp(Z-\eta_0 \rho_0)P_1^\perp=0.
\end{eqnarray}
Consequently $P_1^\perp(Z-\eta_0 \rho_0)P_1^\perp=0$ and $\rho_0 \Sigma^{-1}(Z-\eta_0 \rho_0) \Sigma^{-1} \rho_0=0$ are equivalent propositions for two density matrices without overlapping supports. The same result is of course true when we swap $0$ and $1$. Therefore, 
$P_0^\perp(Z-\eta_1 \rho_1)P_0^\perp=0$ and $\rho_1 \Sigma^{-1}(Z-\eta_1 \rho_1) \Sigma^{-1} \rho_1=0$ are equivalent too.\\

We now come back to the equivalence (\ref{eqbob}). Thanks to the previous discussion, the proposition
\begin{eqnarray}
\exists Z\ge 0 \,\,\,\textrm{such that}\,\,\,
\left\{\begin{array}{c}
ZE_?=0\\
P_0^\perp(Z-\eta_1 \rho_1)P_0^\perp=0\\
P_1^\perp(Z-\eta_0 \rho_0)P_1^\perp=0\\
\end{array}
\right.
\end{eqnarray}
 can be advantageously replaced by
\begin{eqnarray}
\exists Z\ge 0 \,\,\,\textrm{such that}\,\,\,
\left\{\begin{array}{c}
ZE_?=0\\
\rho_1 \Sigma^{-1}(Z-\eta_1 \rho_1) \Sigma^{-1} \rho_1=0\\
\rho_0 \Sigma^{-1}(Z-\eta_0 \rho_0) \Sigma^{-1} \rho_0=0\\
\end{array}
\right.
\end{eqnarray}
or in short
\begin{eqnarray}
\exists Z\ge 0 \,\,\,\textrm{such that}\,\,\,
\left\{\begin{array}{c}
ZE_?=0\\
\rho_i\Sigma ^{-1} Z \Sigma^{-1} \rho_i=\eta_i \rho_i \,\textrm{, for}\,\,\, i=0,1
\end{array}
\right. 
\end{eqnarray}
where we used Eqn.~(\ref{sigma}).\\

\noindent{\bf Third step}\\
Since the operator $Z$ is positive, we know there exists an operator $Y$ such that $Z=Y Y^\dagger$. We can insert it in the relation $\rho_i\Sigma ^{-1} Z \Sigma^{-1} \rho_i=\eta_i \rho_i$ and find, using the decomposition in Eqn.(1), that there exists a unitary transformation $W_i$ such that
\begin{eqnarray}\label{Z2}
W_i^\dagger Y^\dagger \Sigma^{-1}\rho_i=\sqrt{\eta_i}\sqrt{\rho_i},\,\,\, i=0,1.
\end{eqnarray}
Moreover $\Sigma$ is full rank since $\rho_0$ and $\rho_1$ span the total Hilbert space \cite{raynal05a}. We can then decompose $Z$ as $Z=\Sigma \Sigma^{-1} Z \Sigma^{-1} \Sigma=\rho_0 \Sigma^{-1} Z \Sigma^{-1} \rho_0 + \rho_0 \Sigma^{-1} Z \Sigma^{-1} \rho_1 +\rho_1 \Sigma^{-1} Z \Sigma^{-1} \rho_0 + \rho_1 \Sigma^{-1} Z \Sigma^{-1} \rho_1$. This directly yields
\begin{eqnarray}\label{form}
Z&=&\eta_0 \rho_0 + \eta_1 \rho_1 + \sqrt{\eta_0 \eta_1} \sqrt{\rho_0} W_0^\dagger W_1 \sqrt{\rho_1} + \sqrt{\eta_0 \eta_1} \sqrt{\rho_1} W_1^\dagger W_0 \sqrt{\rho_0}\\ \nonumber
&=&(\sqrt{\eta_0} \sqrt{\rho_0} W_0^\dagger W_1 +\sqrt{\eta_1} \sqrt{\rho_1}) (\sqrt{\eta_0}W_1^\dagger W_0 \sqrt{\rho_0} +\sqrt{\eta_1} \sqrt{\rho_1})
\end{eqnarray}
We finally read off $Y^\dagger$ as
\begin{eqnarray}
Y^\dagger=\sqrt{\eta_0}W^\dagger \sqrt{\rho_0} +\sqrt{\eta_1} \sqrt{\rho_1}
\end{eqnarray}
where $W^\dagger=W_1^\dagger W_0.$\\

We now make use of the relation $ZE_?=0$ (Eqn.(\ref{eq1})) which is equivalent to $Y^\dagger E_?=0$. We can explicitly write $Y^\dagger E_?=0$ with $Y^\dagger=\sqrt{\eta_0}W^\dagger \sqrt{\rho_0} +\sqrt{\eta_1} \sqrt{\rho_1}$ and $W=W_0^\dagger W_1$. Therefore the statement:
\begin{eqnarray}\label{bub}
\exists Z\ge 0 \,\,\,\textrm{such that}\,\,\,
\left\{\begin{array}{c}
ZE_?=0\\
\rho_i\Sigma ^{-1} Z \Sigma^{-1} \rho_i=\eta_i \rho_i \,\textrm{, for}\,\,\, i=0,1,
\end{array}
\right.
\end{eqnarray}
can be replaced by:\\

There exists a unitary transformation $W$ such that
\begin{eqnarray}\label{bib}
-\sqrt{\eta_0}W^\dagger \sqrt{\rho_0} E_?=\sqrt{\eta_1} \sqrt{\rho_1}E_?.
\end{eqnarray}
Note that this really is an equivalence and it is not difficult to go from (\ref{bib}) to (\ref{bub}). If a unitary $W$ exists such that $-\sqrt{\eta_0}W^\dagger \sqrt{\rho_0} E_?=\sqrt{\eta_1} \sqrt{\rho_1}E_?$ then we can write $(\sqrt{\eta_0}W^\dagger \sqrt{\rho_0}+\sqrt{\eta_1} \sqrt{\rho_1})E_?=0$ and define the operator $Z=Y Y^\dagger \ge0$ with $Y^\dagger$ as $\sqrt{\eta_0}W^\dagger \sqrt{\rho_0}+\sqrt{\eta_1} \sqrt{\rho_1}$. We then immediately obtain that there exists a positive semi-definite operator $Z$ such that $ZE_?=0$. To recover Eqn.(\ref{bub}) it remains to check that $\rho_i\Sigma ^{-1} Z \Sigma^{-1} \rho_i=\eta_i \rho_i \,\textrm{, for}\,\,\, i=0,1,$ which can be easily done.

Eldar's conditions together with the assumptions that $E_i$ ($i=0,1$) have rank $r_i$ and $\rho_0$ and $\rho_1$ have no overlapping supports are now extremely simplified. We shall now find the final equivalence in two more brief steps.\\

\noindent{\bf Fourth step}\\
We want to prove that:\\

There exists a unitary transformation $-W$ such that $-\sqrt{\eta_0}W^\dagger \sqrt{\rho_0} E_?=\sqrt{\eta_1} \sqrt{\rho_1}E_?$\\

is equivalent to
\begin{eqnarray}
Q^{\mathrm{opt}} = 2\sqrt{\eta_0\eta_1}F.
\end{eqnarray}
It was already proved a very close statement in \cite{raynal05a}:
\begin{eqnarray}
Q^{\mathrm{opt}} = 2\sqrt{\eta_0\eta_1}F \Leftrightarrow
\sqrt{\eta_0}V^\dagger \sqrt{\rho_0} E_?=\sqrt{\eta_1} \sqrt{\rho_1}E_?
\end{eqnarray}
where $V$ is a unitary transformation coming from a polar decomposition of $\sqrt{\rho_0}\sqrt{\rho_1}$.\\

This known result already implies that:
If $Q^{\mathrm{opt}} = 2\sqrt{\eta_0\eta_1}F$, then there exists a unitary transformation $W$ such that $-\sqrt{\eta_0}W^\dagger \sqrt{\rho_0} E_?=\sqrt{\eta_1} \sqrt{\rho_1}E_?$.\\
The other direction is straitforward too. Indeed, from Eldar's equations we notice that $\Tr(Z)=P^{opt}_{success}$ since $\Tr(Z)=\Tr(ZE_?)+\Tr(ZE_0)+\Tr(ZE_1)=\Tr(\sqrt{E_0}Z\sqrt{E_0})+\Tr(\sqrt{E_1}Z\sqrt{E_1})=\Tr(\sqrt{E_0}\eta_0 \rho_0\sqrt{E_0})+\Tr(\sqrt{E_1}\eta_1\rho_1\sqrt{E_1})$ where we used Eqns. (\ref{eq1}), (\ref{eq2}) and (\ref{eq3}) . Therefore, with the form of $Z$ we found in Eqn.~(\ref{form}) (step 3), the optimal failure probability is given by
\begin{eqnarray}
Q^{opt}&=&-\sqrt{\eta_0 \eta_1} ( \Tr(\sqrt{\rho_0} W \sqrt{\rho_1})+ \Tr(\sqrt{\rho_1} W^\dagger \sqrt{\rho_0}) )\\
&=& 2 \sqrt{\eta_0 \eta_1} {\textrm Re} [- \Tr(W^\dagger \sqrt{\rho_0} \sqrt{\rho_1})].
\end{eqnarray}
The fidelity can be expressed as $F=\max_U |\Tr(U^\dagger \sqrt{\rho_0} \sqrt{\rho_1})|$ where the maximum is taken over all the unitary transformations. This already implies that
\begin{eqnarray}
Q^{opt} \le 2 \sqrt{\eta_0 \eta_1}F.
\end{eqnarray}
But of course we know that\cite{rudolph03a,raynal05a}
\begin{eqnarray}
Q^{opt} \ge 2 \sqrt{\eta_0 \eta_1}F.
\end{eqnarray}
In other words, if there exists a unitary transformation $-W$ such that $-\sqrt{\eta_0}W^\dagger \sqrt{\rho_0} E_?=\sqrt{\eta_1} \sqrt{\rho_1}E_?$, then $Q^{opt} = 2 \sqrt{\eta_0 \eta_1}F$ \footnote{Note that we can even conclude that $-W$ comes from a polar decomposition of  $\sqrt{\rho_0}\sqrt{\rho_1}$ since $-\Tr(W^\dagger \sqrt{\rho_0} \sqrt{\rho_1})$ equals $F$.}.\\

\noindent{\bf Fifth step}\\
The proof is almost done. Indeed we only need that
\begin{eqnarray}
Q^{\mathrm{opt}} = 2\sqrt{\eta_0\eta_1}F \Leftrightarrow
\left\{\begin{array}{c}
\rho_0-\sqrt{\frac{\eta_1}{\eta_0}} F_0 \geq 0 \\
\rho_1-\sqrt{\frac{\eta_0}{\eta_1}} F_1 \geq 0
\end{array}
\right.
\end{eqnarray}

however this equivalence has been already proved in~\cite{raynal05a}. This completes the proof. \hfill $\blacksquare$\\

\subsection*{Appendix B: Tighter bounds}

It has been already shown in \cite{herzog05a,raynal05a} that the positivity of the two operators $\rho_0-\sqrt{\frac{\eta_1}{\eta_0}} F_0$ and $\rho_1-\sqrt{\frac{\eta_0}{\eta_1}}F_1$ is only possible when
\begin{eqnarray}\label{reg}
 \frac{\Tr(P_1 \rho_0)}{F}\le \sqrt{\frac{\eta_1}{\eta_0}} \le \frac{F}{\Tr(P_0 \rho_1)}.
\end{eqnarray}
These boundaries were built considering some very general constraints on $Q_0$ and $Q_1$ \cite{raynal05a}:
\begin{eqnarray}\label{cons}
\label{constraintQ0}
\eta_0 \Tr(P_1 \rho_0) \le Q_0 \le \eta_0, \\
\label{constraintQ1}
\eta_1 \Tr(P_0 \rho_1) \le Q_1 \le \eta_1,
\end{eqnarray}
where $P_i$ denotes the projector onto the support of $\rho_i$, $i=0,1$. If more knowledge on the two density matrices $\rho_0$ and $\rho_1$ is provided, we can obtain stronger constraints on $Q_0$ and $Q_1$ and therefore tighter boundaries of the regime (\ref{reg}).\\
Let us give such an example of stronger constraints on $Q_0$ for, say, a POVM having the GU symmetry $E_1=UE_0 U$ where $U^2=\openone$. Since $E_0 \subset {\mathcal K}_{\rho_1}$, there exists $R \ge 0$ in ${\cal K}_{\rho_1}$ such that $P_1^\perp=E_0+R$ and therefore $E_1+E_?=P_1+R$. Moreover the POVM element $E_?$ is invariant under $U$ since $U E_? U =U ({\openone} -E_0 -E_1) U= ({\openone} -E_1 -E_0)=E_?$. Hence, $E_0+E_?=U(E_1+E_?)U = P_0 + URU$. We therefore derive the trace equality
\begin{eqnarray}\label{R}
\Tr(E_?)=2\Tr(R).
\end{eqnarray}
Indeed $\Tr(E_1+E_?)= \Tr(P_1) + \Tr(R)$ and $\Tr(E_0+E_?)= \Tr(P_0) + \Tr(R)$ so that $\Tr({\openone})+\Tr(E_?)=\Tr(P_0)+\Tr(P_1)+2\Tr(R)$. And, for a USD problem in standard form, the equality $\Tr({\openone})=\Tr(P_0)+\Tr(P_1)$ holds.

We can now consider $Q_0$. Since $E_1+E_?=P_1+R$ and $\Tr(E_1 \rho_0)=0$, we can write
\begin{eqnarray}
Q_0&=&\eta_0 \Tr(E_?\rho_0)\\
&=&\eta_0 \Tr(E_? \rho_0) + \eta_0 \Tr(E_1 \rho_0)\\
&=&\eta_0 \Tr(P_1 \rho_0) + \eta_0 \Tr(R \rho_0).
\end{eqnarray}
The operator $P_1^\perp \rho_0 P_1^\perp$ is positive semi-definite. We can here introduce $\lambda_{min}$, its smallest non vanishing eigenvalue. It follows that $Q_0 \ge \eta_0 \Tr(P_1 \rho_0)+ \eta_0 \Tr(R) \lambda_{min}$. Together with Eqn.(\ref{R}) this yields
\begin{eqnarray}
Q_0 &\ge & \eta_0 \Tr(P_1 \rho_0)+ \frac{\eta_0 \lambda_{min}}{2} \Tr(E_?)\\
& \ge & \eta_0 \Tr(P_1 \rho_0)+ \frac{\eta_0 \lambda_{min}}{2} \Tr(E_? \rho_0).
\end{eqnarray}
In other words, for any USD POVM such that $E_1=UE_0 U$ where $U$ is an involution,
\begin{eqnarray}
\frac{\eta_0 \Tr(P_1 \rho_0)}{1-\lambda_{min}/2} \le Q_0
\end{eqnarray}
where $\lambda_{min}=min \{Spect(P_1^\perp \rho_0 P_1^\perp)\}$. This represents a tighter lower bound than the one given in Eqn.(\ref{cons}).

\subsection*{Appendix C: \bf Proof of Corollary~\ref{corollary}}
To prove this corollary we begin with the statement given in Theorem~\ref{rank_theo} for two density matrices $\rho_0$ and $\rho_1$ with the same rank $r$ in a $2r$-dimensional Hilbert space. If the two operators $\rho_0-\sqrt{\frac{\eta_1}{\eta_0}} F_0$ and $\rho_1-\sqrt{\frac{\eta_0}{\eta_1}} F_1$ are not positive semi-definite, Theorem~1 tells us that at least one of the two POVM elements $E_0$ and $E_1$ has rank strictly smaller than $r$. Without loss of generality we say that $r_{E_0}<r$. Because of the completeness relation $E_? +E_1 +E_0= {\openone}$ fulfilled by the POVM elements we have on the support ${\cal S}_{\rho_0}$ the equality $P_0E_?P_0 +P_0E_1P_0 +P_0E_0P_0= P_0$. However ${\cal S}_{E_1} \subset {\cal K}_{\rho_0}$ so that we are left with
\begin{eqnarray}\label{E?}
P_0E_?P_0 +P_0E_0P_0= P_0.
\end{eqnarray}
Furthermore we can consider the spectral decomposition of the Hermitian operator $P_0 E_0 P_0$ and write
\begin{eqnarray}
P_0 E_0 P_0=\sum_{i=1}^{r-1}\lambda_i |\lambda_i \rangle \langle \lambda_i |\\
P_0=\sum_{i=1}^{r-1}|\lambda_i \rangle \langle \lambda_i | + |e \rangle \langle e|
\end{eqnarray}
where $|e \rangle$ completes the $r$ dimensional orthogonal basis of ${\cal S}_{\rho_0}$. As a result $E_? |e \rangle=({\openone} - E_0 -E_1) |e \rangle=|e \rangle - 0 -0$ and $|e \rangle$ is an eigenvector of $E_?$ with eigenvalue $1$. Moreover since $|e \rangle$ is eigenvector with eigenvalue $1$ the completeness relation is already fulfilled onto the subspace spanned by $|e \rangle$. Therefore no optimization is required onto that subspace and we can split it off from the original USD problem. If we denote by ${\cal S}_{|e\rangle}$ the subspace of ${\cal S}_{\rho_0}$ spanned by $|e\rangle$, the reduced Hilbert space is ${\cal H} / {\cal S}_{|e\rangle}$  and the support ${\cal S}_{\rho_0}$ looses one dimension. The remaining USD problem to optimize concerns $\rho_0'$ and $\rho_1'$ originated from the density matrix $\rho_0$ and $\rho_1$. Here $\rho_0'$ has rank $r-1$ while $\rho_1'$ has rank $r$. Thanks to the second reduction theorem, we can reduce this problem to the one of two density matrices of rank $r-1$ in a Hilbert space of dimension $2r-2$. Indeed, the subspace ${\cal K}_{\rho_0'} \cap {\cal S}_{\rho_1'}$ is one dimensional and leads to the detection of $\rho_1'$ with unit probability \cite{raynal03a}. We call $| e' \rangle$ the unit vector spanning this one dimensional subspace. We are left with a reduced USD problem in a $2r-2$ dimensional Hilbert space. Importantly, $| e' \rangle$ is in ${\cal K}_{\rho_0'} \cap {\cal S}_{\rho_1'} \subset {\cal K}_{\rho_0'} = {\cal K}_{\rho_0}$. Indeed, ${\cal H}={\cal S}_{\rho_0} \oplus {\cal K}_{\rho_0} = {\cal S}_{\rho_0'} \oplus {\cal S}_{| e \rangle} \oplus {\cal K}_{\rho_0}$ so that, in ${\cal H'}={\cal H} / {\cal S}_{|e\rangle}$, ${\cal K}_{\rho_0'}={\cal K}_{\rho_0}$.\\
In other words if $\rho_0-\sqrt{\frac{\eta_1}{\eta_0}} F_0$ and $\rho_1-\sqrt{\frac{\eta_0}{\eta_1}} F_1$ are not positive semi-definite, then there exists $|e \rangle$ in ${\cal S}_{\rho_0}$, eigenvector of $E_?$ with eigenvalue $1$, and $| e' \rangle$ in ${\cal K}_{\rho_0}$, eigenvector of $E_1$ with eigenvalue $1$. Without the assumption $r_{E_0}<r_0$ we have the general statement that if $\rho_0-\sqrt{\frac{\eta_1}{\eta_0}} F_0$ and $\rho_1-\sqrt{\frac{\eta_0}{\eta_1}} F_1$ are not positive then there exists $|e \rangle$ in either ${\cal S}_{\rho_0}$ or ${\cal S}_{\rho_1}$, eigenvector of $E_?$ with eigenvalue $1$ and $| e' \rangle$ either in ${\cal K}_{\rho_0}$ and eigenvector of $E_1$ with eigenvalue $1$, or in ${\cal K}_{\rho_1}$ and eigenvector of $E_0$ with eigenvalue $1$. This completes the proof. \hfill $\blacksquare$\\

\subsection*{Appendix D: \bf Proof of statement about the projective measurement}

We want to show that an optimal USD measurement such that $r_{E_0}=r_{E_1}=1$ and $rank(E_?) \le 2$ is necessarily a projective measurement with $rank(E_?)=2$. To do so we can introduce the unit vectors $| x \rangle \in {\cal K}_{\rho_1}$, $| y \rangle \in {\cal K}_{\rho_0}$ and the real numbers $x$ and $y$ in $]0;+\infty[$ (we could in principle restrict $x$ and $y$ to be in $]0;1]$ because probabilities are smaller than $1$) such that
\begin{eqnarray}
E_0= x | x \rangle \langle x| \ge 0,\\
E_1 = y | y \rangle \langle y | \ge 0.
\end{eqnarray}
We call ${\cal S}_{xy}$ the two-dimensional subspace spanned by $| x \rangle$ and $| y \rangle$, $P_{xy}$ the projector onto it and $P_{xy}^\perp$ the projector onto its orthogonal complement. From the definition of the subspace ${\cal S}_{xy}$ and the completeness relation $\sum_k E_k=\openone$ we have
\begin{eqnarray}
P_{xy}^\perp E_? P_{xy}^\perp= P_{xy}^\perp.
\end{eqnarray}
Therefore $rank (P_{xy}^\perp E_? P_{xy}^\perp) = rank (P_{xy}^\perp)=2$ and $E_?$ must be at least of rank $2$. However we already know that $rank(E_?) \le 2$. Therefore $rank (E_?)=2$ and
\begin{eqnarray}
E_?=P_{xy}^\perp.
\end{eqnarray}
We can now consider the subspace ${\cal S}_{xy}$ only. On that subspace, we have
\begin{eqnarray}
E_0 +E_1 =P_{xy}
\end{eqnarray}
that is to say $P_{xy}=x | x \rangle \langle x |+ y | y \rangle \langle y |$. Since $P_{xy}$ is a projector, $P_{xy}=P_{xy}^2$ and it follows that $x^2 | x \rangle \langle x |+ y^2 | y \rangle \langle y | + xy \langle y |x \rangle |y \rangle \langle x| +  xy \langle y |x \rangle |y \rangle \langle x|= x | x \rangle \langle x |+ y | y \rangle \langle y |$. The off-diagonal terms are equal if and only if $\langle y |x \rangle=0$ ($x \neq 0$ and $y\neq 0$) while the diagonal terms are equal if and only if $x=y=1$. Therefore our POVM is a projective measurement with $rank(E_?)=2$.\\

\subsection*{Appendix E: \bf Proof of the statement about the spectrum of $P_1^\perp U P_1^\perp$}

Note that the two operators $P_0^\perp U P_0^\perp$ and $P_1^\perp U P_1^\perp$ have the same spectrum. Therefore, we need to prove the following theorem
\begin{theorem}Spectrum\\
If $\rho_0-F_0$ is not positive semi-definite then $P_0^\perp U P_0^\perp$ has one positive and one negative eigenvalue.
\end{theorem}

First, we write the operator $\rho_0-F_0$ as follows:
\begin{eqnarray}
\rho_0-F_0&=&\rho_0-\sqrt{\sqrt{\rho_0} \rho_1 \sqrt{\rho_0}}\\
&=&\rho_0-\sqrt{\sqrt{\rho_0} U \rho_0 U \sqrt{\rho_0}}\\
&=&\rho_0-\sqrt{\sqrt{\rho_0} U \sqrt{\rho_0} \sqrt{\rho_0} U \sqrt{\rho_0}}\\
&=&\rho_0-|\sqrt{\rho_0} U \sqrt{\rho_0}|
\end{eqnarray}
We now introduce the two orthogonal projectors $P_\pm=\frac{\openone \pm U}{2}$ (since $U^2=\openone$, $P_\pm^2=P_\pm$ and $P_\pm^\dagger=P_\pm$). If $P_0 U P_0$ is positive semi-definite so is $\sqrt{\rho_0} U \sqrt{\rho_0}$ and we simply have $|\sqrt{\rho_0} U \sqrt{\rho_0}|=\sqrt{\rho_0} U \sqrt{\rho_0}$. In that case the operator $\rho_0-F_0$ equals $\sqrt{\rho_0}(\openone - U)\sqrt{\rho_0}=2 \sqrt{\rho_0}P_-\sqrt{\rho_0}$ and is positive semi-definite (since it is of the form $ABA^\dagger$ with $B\ge0$). Similarely, if $P_0 U P_0$ is negative semi-definite then the operator $-\sqrt{\rho_0} U \sqrt{\rho_0}$ is positive semi-definite and we simply have $|\sqrt{\rho_0} U \sqrt{\rho_0}|=-\sqrt{\rho_0} U \sqrt{\rho_0}$. In that case the operator $\rho_0-F_0$ equals $\sqrt{\rho_0}(\openone + U)\sqrt{\rho_0}=2 \sqrt{\rho_0}P_+\sqrt{\rho_0}$ and is positive semi-definite too. The immediate consequence is that if $\rho_0-F_0$ is not positive semi-definite, then $P_0 U P_0$ has one positive and one negative eigenvalue.\\

To complete the proof, we only need the following simple relation between the spectrum of $P_0 U P_0$ and that of $P_0^\perp U P_0^\perp$:
\begin{lemma}
For any two GU states, $Spec(P_0 U P_0)=-Spec(P_0^\perp U P_0^\perp)$.
\end{lemma}
This last statement can be proved in three steps. The first step corresponds to the derivation of the eigenvalues of $P_0+U P_0 U$ from the eigenvalues of $P_0 U P_0$. The second step corresponds to the derivation of the eigenvalues of $P_0+U P_0 U$ from the eigenvalues of $P_0^\perp U P_0^\perp$. In the last step we compare the two different expressions of the eigenvalues of  $P_0+U P_0 U$ previously obtained.\\

First we can see that the eigenvalues and eigenvectors of $P_0 U P_0$ give us the eigenvalues and eigenvectors of $P_0+U P_0 U$. Let us consider $|x\rangle$, an eigenvector of $P_0 U P_0$ with eigenvalue $\lambda$ i.e.\ $P_0 U P_0|x\rangle=\lambda |x\rangle$. Since $S_{P_0 U P_0} \subseteq S_{P_0}$, we also have $P_0|x\rangle=|x\rangle$ and therefore $P_0 U|x\rangle=\lambda |x\rangle$. Let us now consider the operator $P_0+U P_0 U$ together with the vectors $P_\pm |x\rangle$. After some calculations we end up with
\begin{eqnarray}\label{l12}
(P_0+U P_0 U) P_\pm |x\rangle=(1 \pm \lambda)P_\pm  |x\rangle.
\end{eqnarray}
This means that to the eigenvector $|x\rangle$ with eigenvalue $\lambda$ of the operator $P_0 U P_0$ correspond two eigenvectors $P_{\pm}|x\rangle$ with eigenvalues $(1\pm \lambda)$ of the operator $P_0+U P_0 U$.
Following the same idea, one can show that to the eigenvector  $|x^\perp\rangle$ with eigenvalue $\lambda^\perp$ of the operator $P_0^\perp U P_0^\perp$ correspond two eigenvectors $P_{\pm}|x^\perp\rangle$ with eigenvalues $(1\pm \lambda^\perp)$ of the operator $P_0^\perp+U P_0^\perp U$.\\

Next, we link the spectra of the two operators $P_0+U P_0 U$ and $P_0^\perp U P_0^\perp$. One can actually write
\begin{eqnarray}\label{sign}
P_0+U P_0 U&=&(\openone-P_0^\perp)+ U (\openone-P_0^\perp) U\\ \nonumber
&=&2 \openone-(P_0^\perp + U P_0^\perp U).
\end{eqnarray}
This allows us to write the spectrum of the operator $P_0 + U P_0 U$ not only in terms of $\lambda_1$ and $\lambda_2$, the eigenvalues of $P_0 U P_0$ but also in terms of $\lambda_1^\perp$ and $\lambda_2^\perp$, the eigenvalues of $P_0^\perp U P_0^\perp$:
$$Spec(P_0 + U P_0 U)=\{1\pm \lambda_1,1\pm\lambda_2\}=\{1\mp \lambda_1^\perp,1\mp\lambda_2^\perp\}.$$ Note here the important swap between the plus and minus signs.\\

The final step follows the observation that two eigenvectors of $P_0 + U P_0 U$ are in $S_{P_+}$ and two eigenvectors are in $S_{P_-}$. We know that the two eigenvalues corresponding to the two eigenvectors in $S_{P_+}$ are given not only by $1+\lambda_1$ and $1+\lambda_2$ (Eqn. (\ref{l12})) but also by $1-\lambda_1^\perp$ and $1-\lambda_2^\perp$. Indeed from Eqns.(\ref{l12}) and (\ref{sign}), we have for an eigenvector $|x^\perp \rangle$ of $P_0^\perp U P_0^\perp$ with eigenvalue $\lambda^\perp$:
\begin{eqnarray}
(P_0+U P_0 U)P_+|x^\perp \rangle&=&[2 \openone-(P_0^\perp + U P_0^\perp U)]P_+|x^\perp \rangle\\
&=&(1-\lambda^\perp)P_+|x^\perp \rangle.
\end{eqnarray}
 
Since these two pairs of eigenvalues must be identical, we have either
\begin{eqnarray}
1+\lambda_1=1- \lambda_1^\perp\\
1+\lambda_2=1- \lambda_2^\perp
\end{eqnarray}
or 
\begin{eqnarray}
1+\lambda_1=1- \lambda_2^\perp\\
1+\lambda_2=1- \lambda_1^\perp.
\end{eqnarray}
These two cases only differ from their labellings and we finally end up with  $\{\lambda_1,\lambda_2\}=\{-\lambda_1^\perp,-\lambda_2^\perp\}$. In other words, we have obtained that $$Spec(P_0 U P_0)=-Spec(P_0^\perp U P_0^\perp).$$ This completes the proof of the lemma and therefore the proof of the theorem. \hfill $\blacksquare$\\

\bibliography{Bibliography}

\end{document}